August 28, 2023

Interstellar Diplomacy

John Gertz[1,2]

**Abstract**: The Defense Department and NASA are investigating the possibility that aliens are currently surveilling Earth. This aligns with some search-for-extraterrestrial-intelligence (SETI) theorists who have concluded that ET's best strategy for opening a channel of communication is to send artificially intelligent probes to our Solar System for that purpose. This is a golden age for traditional SETI, which is currently well funded, with most of the world's radio telescopes now engaged in the hunt. One way or another, contact with aliens may be imminent. There has been no planning among nations for the aftermath of a first detection. This paper advocates for such planning and diplomacy.

## 1. INTRODUCTION

In the event contact is established between humankind and another technologically competent civilization, the communication between us would create a profound challenge for foreign policy both among nations and between nations and the aliens. Planning for such a contingency should take place now in advance of first contact, rather in the befuddlement of the moment after. There are four drivers for this urgency:

### 1.1 This is the Golden Age of SETI

The search-for-extraterrestrial-intelligence (SETI) has never been better funded. In the United States, Yuri Milner's Breakthrough Foundation has pledged $100M over ten years making its program housed at the University of California, Berkeley and Oxford University, the most robust search ever conducted. Almost every major radio telescope in the world is now engaged in SETI, including China's FAST, the VLA, MEERKAT, the ATA, Parkes Observatory, and Green Bank Observatory among others, as well as some optical telescopes. Between the great increase in the number of telescopes conducting observations and large advances in backend computing power, instrumentation, and detection algorithms it can be estimated that far more SETI is currently conducted every day than was ever before conducted in an entire year. In other words, the chances of the experiment succeeding is at least 1000 times greater than ever.

---

[1] *Zorro Productions, Berkeley, California*
[2] Correspondence address: Zorro Productions, 2249 Fifth St., Berkeley, CA, 94710. Tel: (510) 548-8700. *Email address*: jgertz@zorro.com

August 28, 2023

**1.2 This is the Golden Age of Astronomy**

Quite apart from specific SETI programs, we live in a golden age of astronomy and space science. Almost any telescope or spacecraft might detect evidence for aliens by pure serendipity. For example, a Mars or Moon orbiter might inadvertently photograph an alien artifact, a gamma ray telescope might detect the contrails of an alien matter/antimatter rocket engine, or the James Webb telescope might detect the infrared signature of an alien power plant orbiting a star.

**1.3 ET Probes May Be in our Solar System Surveilling Earth**

The classic SETI paradigm entails the sequential observation of stars for a mere 10 minutes each. Although millions of stars are being surveyed now (including those in the foreground and background of target stars), this paradigm suffers from many problems, chief among which is that the alien transmitter and Earth's receiver must be pointed at each other at the same time (as adjusted for the speed of light). If the aliens are sequentially transmitting to thousands or millions of stars and Earth is likewise listening sequentially to millions of stars, the chance of them aligning in time is negligible [1]. The classic SETI paradigm has been challenged by myself and others who have argued that ET's better strategy for making contact would be to send physical probes to our Solar System for that purpose [2,3,4,5,6,7,8]. We have sent Pioneer and Voyager spacecraft into interstellar space using 1970's technology. An advanced alien civilization can certainly do the same and more. This would yield many advantages to the aliens, among which are:

- An alien artificially intelligent probe might surveil Earth's television and radio transmissions, learning our languages, science, and cultures.
- An alien probe might enter into dialogue with Earth in near real time, rather than with a star-to-star back and forth measured in centuries or millennia.
- It might transmit in a human language it has learned from its surveillance, greatly increasing the chances of mutual comprehension.
- The probe might communicate its findings back to its progenitor civilization. Even before humans had become technologically competent, alien probes could conduct real science, whereas transmissions from the progenitor star yields no information in return unless and until someone on Earth just happens to detect it and Earth transmits a response.
- The alien probe need not reveal the spatial coordinates of its progenitor civilization, thus eliminating any security danger to that civilization. A transmission from its home star by definition reveals its coordinates.
- The alien probe might assess any potential danger posed by Earth's life, providing its progenitor civilization with interstellar situational awareness.

**1.4 ET Probes May Even be in Earth's Atmosphere.**

There has been broad interest in Unidentified Anomalous Phenomena (UAPs), one or more of which may prove to be of alien derivation. The U.S. Air Force has recently upgraded its investigation with the August 4, 2020 announcement of the establishment within the Pentagon of an UAP Task Force created by Deputy Defense Secretary David Norquist. The Department of Defense now has an "All-domain Anomaly Resolution Office" (commonly called AARO), under the current directorship of Sean Kirkpatrick.  On October 21, 2022, NASA announced the creation of its own blue-ribbon panel to study the phenomenon. Purported sightings by air force pilots of objects that defy all known aerodynamics in their movements leads to the deduction that if the observed phenomena really are objects of alien derivation, they must be robotic probes rather than "manned" by little green men. This is because biological beings would presumably be crushed by the G-forces of their very large purported accelerations [9].

2.     **QUESTIONS WITHOUT CURRENT ANSWERS?**

What happens next if humankind actually does make contact with aliens?  The short answer is that nobody knows.  There has been no known (i.e., unclassified) planning.

**2.1  The Detection of an Interstellar Signal**

Let's take first the situation where a transmission is detected from a faraway star per the classic SETI paradigm.  The message may be completely indecipherable, at least at first. It might need to be studied for decades or more by cryptologists, linguists and others to be properly decoded, if at all. It may prove to be ambiguous.  For example, "our only wish is to serve man" might be a manifesto of altruistic intentions or a cookbook. It may be only a simple ping, either devoid of any content or too weak to discern the content embedded within a carrier wave, but surely interpretable as an invitation to signal back to indicate that we also exist as technologically-competent beings, have heard their message and are ready to start a dialogue. In that case, the initial decision is a deceptively simple one, should we send a response or not?  That decision needs to be made now in advance of the receipt of the message because we have no idea how long the alien receiver will dwell on Earth before it moves on to its next target.  Will its receiver dwell on Earth for a few minutes, a month, a year, a century before it concludes that either no technological civilization has received its message or has decided not to respond?   If humankind makes the decision now that we will transmit, then we can do so immediately upon the detection of an alien signal.  We might decide to respond with a simple ping devoid of content, but implying "we hear you, hold on while we try to decipher your message," or maybe add something benign like a string of prime numbers to suggest we have some intelligence. Perhaps we might decide to send a more elaborate message.  Whatever our collective decision, now is the time to make it.

**2.1.1 Should we respond at all?**

However, the decision as to whether to send our own message of predetermined detail or to remain silent is fraught. The answer is not an obvious "yes." The received transmission might be from a hostile civilization, who, like a bat using echolocation to catch its prey, is looking for a bounce back signal from some naïve and new civilization in order to destroy it.  To believe that aliens could not possibly destroy us from astronomical distances may be as naïve as the Assyrians imagining that warfare is impossible beyond the distance an arrow might fly. They would fail to imagine artillery and jet bombers, much less nuclear tipped missiles. One very plausible reason why we do not hear a cacophony of signals from innumerable alien civilizations is that they know something we do not, namely, that the galaxy contains some very bad actors.

We do not know whether aliens would seek to aggress, assist, proselytize, trade (presumably data is the coin of the realm), or simply lurk and surveil without offering anything in return. Whether aliens more closely resemble the loveable "ET" of Steven Spielberg or the fearsome "Aliens" of Ridley Scott is purely a matter of personal taste and conjecture at this point. The one thing, though, that we do know about any aliens whom we might encounter is that they are far more advanced than ourselves. 95% of all stars are older than our Sun. With billions of years in which they may have evolved before us, the chances that they too are only in their first century of being able to send and receive signals or launch interstellar probes is statistically negligible.

### 2.1.2 Encumbering Future Generations

If a message has been received from the distance of, say, 1000 light years, and we choose to respond, then ET's response to us will only be received 2000 years hence. We thus encumber unborn generations with our decision today. That ethical problem is compounded if there is no agreement among nations on whether to return the alien's signal or not. As matters stand now, any one country might make the decision to signal by itself, and thereby encumber all humankind and its future generations.

### 2.1.3 What Information Should Be Released to the Public?

There is no valid reason to hide the fact of an alien detection from the public. If the Chinese or Americans are the first to detect a signal coming from a distant star, will they share the coordinates with each other or mark it a state's secret so that they alone might benefit? Perhaps the coordinates should be placed into the neutral hands of the United Nations Secretary General. In order to deter an unauthorized response, perhaps from a rogue nation or a religious cult, the Secretary General might be empowered to keep the coordinates secret from the general public.

### 2.1.4 Alien Probes Within Our Own Solar System

What might be the plan if an alien probe is detected, perhaps as close as the Moon or closer yet? People might not panic if aliens are detected a thousand light years away. But they might if the aliens are detected at the distance of Venus or even the Moon, mere light seconds or minutes from Earth. It is unlikely that rational decisions would be taken in the moment, rather than with forethought, and with all nations acting in concert.

Should we shoot the probe down, or invite it to land at Buckingham Palace? Should the probe be treated as an ambassador and be given diplomatic immunity even though it is obviously nonbiological? Perhaps ET's message will be seemingly benign, with an offer to immediately download for our benefit the great and grand *Encyclopedia Galactica*. But could this be a Trojan horse ready to infect our entire Internet? Could this artificial genie give us blueprints for little nano-robots that once injected into the blood stream cure cancer, but that might also do who knows what other things to more nefarious ends [10]?

What if an alien probe threatens us along the lines of "we have seen your wretched nightly news, want nothing to do with you, and have marked you for destruction." Is the probe bluffing? Should we attempt to shoot the probe down then? Would the probe issue such a threat if it could not in fact thwart our attack? Would it actually warn us if it was seriously intent on our destruction? Is the probe merely trying to test us, or to jolt us into better behavior? If we opt to shoot it down, might we inadvertently trigger the very destruction we are trying to avoid? Should the nations of Earth act in coordination, or will it be everyone for themselves?

What if some nation or private party comes into physical possession of a probe? It could happen if, for example, a space mining company of the near future encounters a probe on an asteroid, or if the probe lands on the proverbial White House lawn, or is in orbit around the Moon and one nation sends a spacecraft to retrieve it? Who then owns the probe? In whose laboratory is it to be studied? Or is it not to be studied at all but treated as an ambassador?

If a probe's laser transmission is narrowly focused on the Keck observatory in Hawaii, allowing the U.S. alone the ability to receive its message, will the Chinese be incentivized to either shoot down the probe or bomb Hawaii lest the U.S. receive a cornucopia of technologies that would highly disadvantage them? Perhaps UN observers should be assigned to each nation's SETI program to prevent unilateral actions or advantages.

If we detect a probe before it actively transmits to us (perhaps a small "asteroid" is discovered to be artificial), should we transmit a greeting to it or leave it undisturbed until it is ready to transmit to Earth? If we do not initiate a transmission, then we risk the possibility that it may be millennia, if ever, before that probe deigns to communicate with us. Even more, it may well be that the galactic protocol requires us to transmit first if we want to communicate—that is, giving us the choice as to whether we wish to make contact. On the other hand, instead of granting us three wishes like an awakened genie, perhaps our greeting will provoke a defensive response of unpredictable but potentially catastrophic consequences.

### 3.     ANSWERS MIGHT DERIVE FROM PROCESS

There are no easy decisions. This is why we should all be in this together and make these tough choices through representative bodies and codify those decisions within an international treaty.

There should be a multidisciplinary and eminent committee and reporting structure in place, presumably to the UN, that might spend years or decades in thinking out, debating and planning for as many contingencies as it can imagine. Members might include experts in such fields as astrobiology, astronomy, biology, computer science, cryptology, diplomacy, economics, emergency planning, epidemiology, game theory, law, linguistics, mathematics, psychology, religion, rocket science, security, and space science (intentionally listed alphabetically so as not to suggest an order of importance).

### 3.1 COPUOS

It could be that such a committee actually already exists, the United Nations Committee for the Peaceful Use of Space (COPUOS), which was created in 1957 and which has the responsibility for recommending treaties for consideration by the United Nations. The Treaty on Principles Governing Activities of States in the Exploration and Use of Outer Space (the "Outer Space Treaty") was largely crafted by the U.S. and the Soviet Union and recommended by COPUOS in 1967 [11]. It has been signed or adopted by all major countries.

However, COPOUS may not be in the best position to recommend a new treaty governing an alien encounter. It meets infrequently, works by consensus (which is cumbersome), and does not currently have among its membership the full range of expertise needed to advise on alien encounters. Moreover, the Outer Space Treaty it recommended contains no enforcement mechanism or stated penalties. This was made apparent when the treaty was grossly violated by the Chinese in 2007, when they intentionally destroyed one of their own weather satellites with a missile, creating thousands of shards of space junk, and in the process jeopardized hundreds or thousands of low earth orbit satellites. China's reckless action was met with broad condemnation, but no punitive consequences. A SETI Treaty will need teeth.

### 3.2 Alternatives to COPUOUS

There are alternatives to COPUOS and the UN. Recognizing that there are very few countries with serious SETI and astronomical capabilities, it is possible that a multilateral treaty, outside of the framework of the United Nations, may suffice. Perhaps a simple bilateral treaty between China and the U.S. might serve as a beginning, with other SETI, space and astronomically advanced countries joining later.

### 3.3 Transparency and Inspections

The envisioned treaty should contain provisions for inspections and verification. Every SETI, space and astronomical program should be open to all signatories for intrusive inspection. Chinese scientists should have the right to receive American data streams and vis-a-versa. Currently, all U.S. SETI data is open to the Chinese, while the opposite is not true. There is a real possibility that the Chinese would not admit to a detection without this type of cross-inspection. American SETI scientists are pledged to transparency, but one leading researcher confided in me that he fully expects that "men in black suits" will take him away the moment he might make the first detection.

### 4. A FIRST DRAFT TREATY NEED NOT BE ROCKET SCIENCE

It does not require a large committee and years of debate to put forth at least a first draft of a proposed treaty. I therefore set forth a potential first draft of a treaty governing alien contact in an effort to answer as many of the questions I have raised as possible. I have adopted or adapted some of the language in The Outer Space Treaty, and added notes in brackets.

August 28, 2023

# Treaty on Principles Governing the Activities of States in Humankind's Relations with Robotic or Biological Extraterrestrial Intelligence.

The States Parties to this Treaty,

Inspired by the great prospects of discovering evidence for and communicating with Robotic or Biological Extraterrestrial Intelligence ("Alien Beings"),

Recognizing the common interest of all humankind in establishing peaceful relations with such Alien Beings,

Believing that relations with Alien Beings should be carried on for the benefit of all peoples irrespective of the degree of their economic or scientific development,

Believing that all relations with Alien Beings should be carried out by representatives of and on behalf of all humankind,

Understanding that we currently know nothing of the prevalence, nature, intention, or capabilities of such Alien Beings that we might encounter,

Understanding that a first discovery of Alien Beings might derive from intercepted interstellar radio, optical or other electromagnetic messages, or might derive from evidence gathered independently of any attempt on the part of aliens to communicate with humankind (i.e., the discovery of their technosignatures), or might derive from the recording of alien signals originating from within our own Solar System, or within Earth's own atmosphere or biosphere,

Recalling the Treaty on Principles Governing the Activities of States in the Exploration and Use of Outer Space, including the Moon and Other Celestial Bodies,

Convinced that a Treaty on Principles Governing the Activities of States in Humankind's Relations with Robotic or Biological Extraterrestrial Intelligence will further the purposes and principles of the Charter of the United Nations,

Have agreed on the following:

<div style="text-align:center">Article I</div>

All States Parties shall be entitled to conduct searches for extraterrestrial intelligences ("SETI"). All such SETI searches shall be conducted for the benefit of all humankind.

<div style="text-align:center">Article II</div>

States Parties to the Treaty shall immediately inform all other States Parties to the Treaty or the Secretary-General of the United Nations of any evidence of Alien Beings, including apparently biological beings, apparently artificial beings, apparently robotic alien devices, apparently inert alien artifacts, and apparent alien technosignatures [a technosignature is any evidence of intelligent aliens, such as a physical artifact, artificial signatures in the atmosphere of an exoplanet, a large obviously non-natural object detected in orbit around an exoplanet or star, the gamma ray contrails of an interstellar space craft, etc.] of any type.

Article III

States Parties to the Treaty shall bear international responsibility for national SETI activities, whether such activities are carried on by governmental agencies or by non-governmental entities, and whether the SETI activity is direct or tangential to normative astronomical and space research, and for assuring that national SETI activities are carried out in conformity with the provisions set forth in the present Treaty. When SETI activities are carried out by an international organization, responsibility for compliance with this Treaty shall be borne both by the international organization and by the States Parties to the Treaty participating in such organization. The provisions of this Treaty shall apply to the activities of States Parties to the Treaty, whether such activities are carried out by a single State Party to the Treaty or jointly with other States, including cases where they are carried out within the framework of international intergovernmental organizations.

Article IV

The Secretary General of the United Nations in consultation with the States Parties to this Treaty shall create and maintain a Standing Committee of Experts ("SCE") in diverse relevant fields, including, without limitation, astrobiology, astronomy, biology, computer science, cryptology, diplomacy, economics, emergency planning, epidemiology, game theory, law, linguistics, mathematics, psychology, religion, rocket science, security, and space science [intentionally listed alphabetically so as not to suggest an order of importance]. From the larger body of the SCE, the Secretary General shall appoint an Executive Committee, which, in turn, shall appoint a Director. The SCE shall have among its duties the confirmation of a detected alien signal, decryption of an alien message, the construction of Earth's authorized response, if any, and the study of any physical alien artifact. The SCE shall debate internally and advise the Secretary General on strategies for a response to the detection of intelligent biological or artificial Alien Beings, robotic alien probes whether active or apparently dormant, or technosignatures of any kind or nature. The SCE shall be headquartered in Geneva, Switzerland, and be funded by the States Parties to this Treaty in proportions and with a budget as apportioned among the States Parties as determined by the Secretary General.

Article V

The fact of an alien detection shall be made public by the Secretary General of the United Nations in coordination with the party that made the first discovery. Such announcement shall proceed as soon as feasible after the confirmation of the detection. However, the coordinates of the detection as well as orbital characteristics if any and the content of any message, whether decipherable or not, as well as the frequency of an alien transmission if any shall be conveyed only to the Secretary General of the United Nations who, in consultation with the Executive Committee of the SCE, shall have the right to redact such information as he or she sees fit and to constrain the SCE and all other relevant parties to secrecy to the extent deemed necessary.

## Article VI

The Secretary General of the United Nations shall have the right to order All States to discontinue the use of any frequency used by aliens to communicate with Earth, except only that such frequency may be used by the United Nations for the purpose of communicating with aliens.

## Article VII

No response to an alien transmission shall be made without the expressed approval of three-quarters of the UN Security Council including not less than four of its permanent members, as well as three-quarters of the member States of the United Nations General Assembly.

## Article VIII

All SETI programs, installations, and equipment shall be open to representatives of other States Parties to the Treaty. Such representatives shall give reasonable advance notice of a projected visit, in order that appropriate consultations may be held and that maximum precautions may be taken to assure safety and to avoid interference with normal operations of the facility to be visited. Any or all raw data streams or reductions thereof from all SETI observations shall be made available in real time to all other SETI researchers upon request.

## Article IX

Any physical alien object of any sort coming into the possession of any State Party or any party within its jurisdiction shall become the property of all humankind and shall be administered by the United Nations. If discovered in space, if at all feasible, it shall be studied *in situ* and not brought to Earth without the explicit vote of the United Nations Security Council as advised by the Executive Committee of the SCE. In the event that the alien object appears to be either sentient or to harbor sentient beings, and unless deemed hostile in its intent, per Article X below, it shall be treated as an alien ambassador and accorded all rights accorded to States ambassadors. No suspected alien probe whether located in space or within Earth's atmosphere, waters, or on land may be harmed in any way unless it is deemed hostile in accordance with Article X below.

## Article X

The right to declare aliens or their technology as hostile shall reside solely with the UN Security Council, subject to the right of each State to defend itself against an ongoing or imminent attack. In the event that the aliens are determined to be hostile by vote of the UN Security Council, command of Earth's defensive response shall reside with the United Nations.

## Article XI

The UN Security Council, as advised by the Executive Committee of the SCE, shall be empowered to adopt sanctions or such other actions or penalties as it deems appropriate against nation states or other individuals or entities that violate the terms of this Treaty. In the event that the nation state that may be the subject of such sanctions is a member of the Security Council, it shall not be allowed a vote or a veto with reference to such sanctions or other actions or penalties.

## Article XII

Transmitting to aliens in advance of their detection, so-called messaging-to-extraterrestrial-intelligence ("METI"), is expressly prohibited [12; METI is the proactive signaling of stars, hoping to receive a response.  It may be necessary to further define METI as directed transmissions above defined power thresholds for each electromagnetic wavelength. Obviously, there is no interest in preventing a child from using a flashlight to signal stars]. States Parties to the Treaty shall enact criminal laws against METI with penalties of not less than five (5) years, nor more than thirty (30) years of incarceration. Planetary radar shall be allowed for the purpose of asteroid threat assessment. However, this shall be subject to the prior approval of the SCE Executive Committee and shall not target celestial objects that are occulting nearby stars or within fifteen (15) degrees of the midline of the plane of the Milky Way [where the density of stars is greatest].

## Article XIII

1. This Treaty shall be open to all States for signature. Any State that does not sign this Treaty before its entry into force in accordance with Paragraph 3 below of this article may accede to it at any time.

2. This Treaty shall be subject to ratification by signatory States. Instruments of ratification and instruments of accession shall be deposited with the Governments of the South Africa, the People's Republic of China, and the United States of America, which are hereby designated the Depositary Governments.  [The three depository nations are chosen to represent East, West and South. All three nations currently engage in SETI, while the largest radio telescope in the world, the Square Kilometer Array (SKA), is currently under construction in South Africa and expected to be complete in 2028. It will perform SETI observations as well as normative astronomy.]

3. This Treaty shall enter into force upon the deposit of instruments of ratification by five Governments including the Governments designated as Depositary Governments under this Treaty.

4. For States whose instruments of ratification or accession are deposited subsequent to the entry into force of this Treaty, it shall enter into force on the date of the deposit of their instruments of ratification or accession.

5. The Depositary Governments shall promptly inform all signatory and acceding States of the date of each signature, the date of deposit of each instrument of ratification of and accession to this Treaty, the date of its entry into force and other notices.

6. This Treaty shall be registered by the Depositary Governments pursuant to Article 102 of the Charter of the United Nations.

## Article XIV

Any State Party to the Treaty may propose amendments to this Treaty. Amendments shall enter into force for each State Party to the Treaty accepting the amendments upon their acceptance by a majority of the States Parties to the Treaty and thereafter for each remaining State Party to the Treaty on the date of acceptance by it.

Article XV

Any State Party to the Treaty may give notice of its withdrawal from the Treaty one year after its entry into force by written notification to the Depositary Governments. Such withdrawal shall take effect one year from the date of receipt of this notification.

Article XVI

This Treaty, of which the English, Swahili, French, Spanish, and Chinese texts are equally authentic, shall be deposited in the archives of the Depositary Governments. Duly certified copies of this Treaty shall be transmitted by the Depositary Governments to the Governments of the signatory and acceding States.

IN WITNESS WHEREOF the undersigned, duly authorized, have signed this Treaty.

DONE in triplicate, at the cities of Cape Town, Beijing, and Washington, D.C., this ____ day of _________, in the year ______.

## 5. CONCLUSIONS

We do not know whether alien civilizations that seek to communicate with Earth exist. However, if they do, then we might hear from one or more in the near future. This may be because we are engaged in a more robust SETI program than ever before; or because this is a golden age of astronomical discovery and an ET detection has occurred serendipitously; or it may be because a very nearby civilization has detected our omnidirectional electro-magnetic leakage; or it may be from a local alien probe that has decoded that leakage and is able to communicate in a learned terrestrial language. It behooves humankind to prepare for such an eventuality by appointing a committee of multi-disciplinary experts to study and debate the multifarious contingencies and recommend humankind's response. This activity should be organized and governed by principles laid out in an international treaty.

## 6. ACKNOWLEDGEMENTS

The author gratefully acknowledges useful suggestions from Ivo Keller and JBIS's two reviewers of this paper.

August 28, 2023